\documentclass{emulateapj}
\usepackage{natbib} 
\begin{document}
\def\mpch {$h^{-1}$ Mpc} 
\def\kms {km s$^{-1}$} 
\def\lcdm {$\Lambda$CDM } 
\def\xir {$\xi(r)$}
\def\wprp {$w_p(r_p)$}
\def\xisp {$\xi(r_p,\pi)$}
\def\rr {$r_0$}

\title{The DEEP2 Galaxy Redshift Survey: Clustering of Galaxies as a
  Function of Luminosity at $z=1$}
\author{Alison L. Coil\altaffilmark{1,2}, 
Jeffrey A. Newman\altaffilmark{1,3},
Michael C. Cooper\altaffilmark{4}, 
Marc Davis\altaffilmark{4},
S.~M. Faber\altaffilmark{5},
David C. Koo\altaffilmark{5},
Christopher N.~A. Willmer\altaffilmark{2,5}
}
\altaffiltext{1}{Hubble Fellow}
\altaffiltext{2}{Steward Observatory, University of Arizona,
Tucson, AZ 85721}
\altaffiltext{3}{Institute for Nuclear and Particle Astrophysics,
  Lawrence Berkeley National Laboratory, Berkeley, CA 94720} 
\altaffiltext{4}{Department of Astronomy, University of California,
Berkeley, CA 94720}
\altaffiltext{5}{University of California Observatories/Lick
Observatory, Department of Astronomy and Astrophysics, University of
California, Santa Cruz, CA 95064}

\begin{abstract}

We measure the clustering of DEEP2 galaxies at $z=1$ as a function of
luminosity on scales $0.1$ \mpch \ to $20$ \mpch.  Drawing
from a parent catalog of 25,000 galaxies at $0.7<z<1.3$ in the
full DEEP2 survey, we create volume-limited samples having upper
luminosity limits between $M_B=-19$ and $M_B=-20.5$, roughly $0.2-1
L^*$ at $z=1$.  We find that brighter galaxies are more strongly
clustered than fainter galaxies and that the slope of the correlation
function does not depend on luminosity for $L<L^*$.  The brightest
galaxies, with $L>L^*$, have a steeper slope.  The clustering
scale-length, \rr, varies from $3.69 \pm0.14$ for the faintest sample 
to $4.43 \pm0.14$ for the brightest sample.  The relative bias of galaxies as
a function of $L/L^*$ is steeper than the relation found locally for SDSS
galaxies \citep{Zehavi05} over the luminosity range that we sample.
The absolute bias of galaxies at $z\sim1$ is scale-dependent on scales
$r_p<1$ \mpch, and rises most significantly on small scales for the
brightest samples. For a concordance cosmology, the large-scale bias
varies from $1.26 \pm0.04$ to $1.54 \pm0.05$ as a function of
luminosity and implies that DEEP2 galaxies reside in dark matter halos
with a minimum mass of $\sim1-3 \times10^{12} h^{-1} {\rm M}_\sun$.

\end{abstract}

\keywords{galaxies: high-redshift --- cosmology: large-scale structure 
of the universe}

\section{Introduction}

The clustering of galaxies has long been used as a fundamental measure
of the large-scale structure of the universe.  Clustering measures
place strong constraints on galaxy formation and evolution models and
provide estimates of the average or minimum parent dark matter halo
mass of a given galaxy population, allowing placement in a
cosmological context \citep[e.g.,][]{Mo96, Sheth99}.  Locally, large
surveys such as the 2-Degree Field Galaxy Redshift Survey (2dFGRS) and
the Sloan Digital Sky Survey (SDSS) have measured the two-point
correlation function of galaxies precisely enough to allow detailed
halo occupation distribution (HOD) modeling \citep[e.g.,][]{Yan03b,
Phleps05,Yang05HOD,Zehavi05}, as well as cosmological parameter
constraints \citep[e.g.,][]{Peacock01,Abazajian05}.

In this paper we focus on the observed clustering of galaxies as a
function of luminosity. Both 2dF and SDSS have found that at
$z\sim0.1$ brighter galaxies cluster more strongly, with the relative
bias on large scales increasing linearly with luminosity
\citep[e.g.,][]{Alimi88,Hamilton88,Benoist96, Willmer98,Norberg01,Norberg02,Zehavi04}.  A
simple interpretation is that brighter galaxies reside in more massive
dark matter halos, which are more clustered in the standard theories
of hierarchical structure formation.  The full galaxy samples for
these surveys are generally unbiased with respect to the dark matter
density field on large scales
\citep[e.g.,][]{Peacock01,Verde01,Croton04}.

Locally, the two-point correlation function is relatively well-fit by
a power law, \xir=$(r/r_0)^{-\gamma}$, with $r_0\sim5$ \mpch \ and
$\gamma\sim1.8$ \citep[e.g.,][]{Norberg02,Zehavi05}.  Small deviations
from a power law have recently been detected on scales $r<10$ \mpch \
and can be naturally be explained in the HOD framework as the
transition between pairs of galaxies within the same halo (the
`one-halo' term) on small scales and galaxies in different halos (the
`two-halo' term) on larger scales \citep{Zehavi04}.  The largest
deviations are seen for the brightest galaxy samples, with $L>L^*$.
These galaxies are generally found in groups and clusters, which have
larger `one-halo' terms due to the many galaxies residing in a single
parent dark matter halo and therefore have a steeper correlation slope
on small scales.

Lyman-break galaxies (LBGs) at $z\sim3-4$ also display a
luminosity-dependence in their clustering properties
\citep{Allen05,Ouchi05,Lee05}.  The large-scale bias of LBGs ranges
from $\sim3-8$, depending on the luminosity of the sample. There is
also a strong scale-dependence to the bias on small scales, where the
correlation function rises sharply below $r\sim0.5$ \mpch
\citep{Ouchi05, Lee05}.  This rise has been interpreted as evidence
for multiple LBGs residing in single dark matter halos at $z\sim4$.

In this paper, we present results on the luminosity-dependent
clustering of galaxies at $z\sim1$, using the nearly-completed DEEP2
Galaxy Redshift Survey.  The data sample used here is 10 times larger
than our initial results presented in \cite{Coil03xisp} and represents
the most robust measurements of \xir \ at $z>0.7$ to date.  In future
papers we will address clustering properties as a function of other
galaxy properties such as color, stellar mass and redshift; here we
focus solely on the luminosity-dependence.  We choose to use integral,
rather than differential, luminosity bins to facilitate comparison
with theoretical models \citep[e.g.,][]{Kravtsov04,Tinker05}. 
Luminosity-threshold samples are preferred for HOD modeling, as 
theoretical predictions for HODs have been studied more extensively 
for mass (luminosity) thresholds and afford fewer parameters to fit 
than differential bins \citep[for details see e.g.][]{Zehavi05}.  
In this case the HOD has a relatively simple form, e.g., the mean 
occupation function is a step function (for central galaxies) plus 
a power law (for satellites).  The
outline of the paper is as follows: \S 2 briefly describes the DEEP2
survey and data sample.  In \S 3 we outline the methods used in this
paper, while \S 4 presents our results on the luminosity-dependence of
galaxy clustering at $z\sim1$.  In \S 5 we discuss the galaxy bias and
the relative biases between our luminosity samples and conclude in \S
6.

\section{Data Sample}

\begin{figure}[t]
\centerline{\scalebox{0.4}{\includegraphics{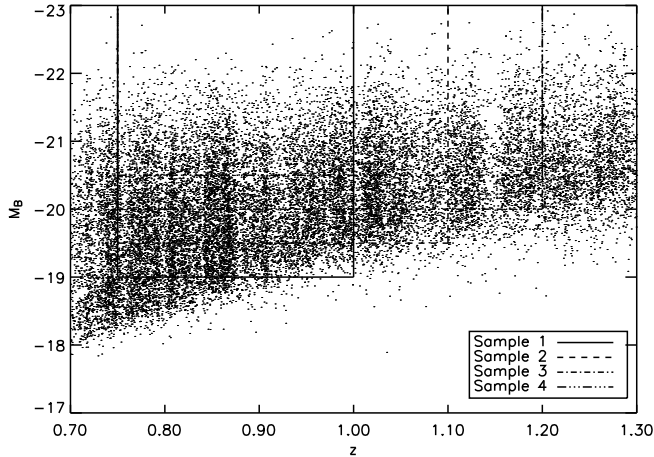}}}
\caption{Absolute B-band magnitude (in AB magnitudes, with $h=1$) 
versus redshift for the parent DEEP2 galaxy catalog. The magnitude and 
redshift ranges of the four samples used here are shown, and properties of
each sample are given in Table 1.
\label{magz}}
\end{figure}

The parent galaxy sample used here is from the DEEP2 Galaxy Redshift
Survey, a three-year project using the DEIMOS spectrograph
\citep{Faber03} on the 10-m Keck II telescope to survey optical
galaxies at $z\simeq1$ in a comoving volume of approximately
5$\times$10$^6$ $h^{-3}$ Mpc$^3$.  Using $\sim1$ hour exposure times,
the survey has measured redshifts for $\sim30,000$ galaxies in the
redshift range $0.7\sim z\sim1.45$ to a limiting magnitude of $R_{\rm
AB}=24.1$ \citep{Davis03, Faber06}.  The survey covers three square
degrees of the sky over four widely separated fields to limit the
impact of cosmic variance.  Fields consist of two to four adjacent
pointings of size $0.5\times0.67$ degrees each.  
Details of the DEEP2 observations, catalog construction and data
reduction can be found in Davis et al. (2003), Coil et al. (2004b),
Davis, Gerke, \& Newman (2004), and Faber et
al. (2006)\nocite{Davis03, Coil04, Davis04, Faber06}.  Here we use the
full sample from ten pointings in all four fields to create four
nearly volume-limited samples as a function of Johnson B absolute magnitude
($M_B$). Fig. \ref{magz} shows the absolute magnitude and redshift of
each galaxy in the parent catalog and the regions from which the
different luminosity samples were drawn.  The three brightest samples
(with limiting magnitudes down to $M_B=-19.5$) are volume-limited for
blue galaxies,
while the faintest sample is missing some fainter blue galaxies at
$z\gtrsim0.95$.
The samples are not entirely
volume-limited for red galaxies, due to the $R_{\rm AB}$ selection of the 
DEEP2 survey (see \cite{Willmer06} for definitions of blue and red
galaxies and a discussion of this selection effect).  
  K-corrections are calculated as described in
\cite{Willmer06}; we do not include corrections for luminosity
evolution.  Table \ref{subtable} lists the details of each luminosity
sample.  To convert measured redshifts to comoving distances along the
line of sight we assume a flat \lcdm cosmology with $\Omega_{\rm
m}=0.3$ and $\Omega_{\Lambda}=0.7$.  We define $h \equiv {\rm {\it
H}_0/(100 \ km \ s^{-1} \ Mpc^{-1}})$ and quote correlation lengths,
\rr, in comoving \mpch.  All absolute magnitudes are AB with $h=1$.

\begin{deluxetable}{crccccccc}
\tablewidth{0pt}
\tablecaption{Details of DEEP2 Galaxy Luminosity Samples}
\tablehead{
\colhead{}&\colhead{No. of}  &\colhead{n} &\colhead{$M_B$}&\colhead{Median}&\colhead{$z$}  &\colhead{Mean}\\
\colhead{} &\colhead{galaxies}&\colhead{($h^3 Mpc^{-3}$)}&\colhead{range}&\colhead{$M_B$} &\colhead{range}&\colhead{$z$}             
}
\startdata
1     &  10530  & $1.3 \ 10^{-2}$ & $\le -19.0$   &  $-20.28$   & $0.75-1.0$  & $0.87$   \\
2     &  11023  & $8.4 \ 10^{-3}$ & $\le -19.5$   &  $-20.41$   & $0.75-1.1$  & $0.92$   \\
3     &   9526  & $4.9 \ 10^{-3}$ & $\le -20.0$   &  $-20.64$   & $0.75-1.2$  & $0.98$   \\
4     &   5418  & $2.5 \ 10^{-3}$ & $\le -20.5$   &  $-20.96$   & $0.75-1.2$  & $0.99$   \\
\enddata
\label{subtable}
\end{deluxetable}

\section{Methods}

Details of the methods used to measure the correlation function are
given in \cite{Coil03xisp, Coil05}; we repeat the most relevant
details here.  The two-point correlation function \xir \ is defined 
as a measure of the excess probability above Poisson of finding an 
object in a volume element $dV$ at a separation $r$ from another 
randomly chosen object,
\begin{equation}
dP = n [1+\xi(r)] dV,
\end{equation}
where $n$ is the mean number density of the object in question 
\citep{Peebles80}.  

We measure the two-point correlation function using the
\citet{Landy93} estimator,
\begin{equation}
\xi=\frac{1}{RR}\left[DD \left(\frac{n_R}{n_D}
\right)^2-2DR\left(\frac{n_R}{n_D} \right)+RR\right],
\end{equation}
where $DD, DR$, and $RR$ are pair counts of galaxies in the data-data,
data-random, and random-random catalogs, and $n_D$
and $n_R$ are the mean number densities of galaxies in the data and
random catalogs.  The random catalog has the same overall sky coverage
and redshift distribution as the data and serves as an unclustered
sample with which to compare the data.  The spatial window function of
the DEEP2 survey is applied to the random catalog, which includes
masking areas around bright stars and taking into account the varying
redshift completeness of our observed slitmasks.  

Distortions in $\xi$ are introduced parallel to the line of sight due
to peculiar velocities of galaxies.  In order to uncover the
real-space clustering properties, we measure $\xi$ in two dimensions, 
as a function of distance both perpendicular to ($r_p$) and along 
($\pi$) the line of sight. 
As redshift-space distortions affect only the line-of-sight component
of $\xi$, integrating over the $\pi$ direction leads to the projected
correlation function \wprp, which is independent of redshift-space 
distortions.  Following \cite{Davis83},
\begin{equation}
w_p(r_p)=2 \int_{0}^{\infty} d\pi \ \xi(r_p,\pi)=2 \int_{0}^{\infty}
dy \ \xi(r_p^2+y^2)^{1/2},
\label{eqn}
\end{equation}
where $y$ is the real-space separation along the line of sight. 
We sum $\xi(r_p,\pi)$ to a maximum separation of $\pi=20$ \mpch \ and 
measure \wprp \ as a function of scale for $0.1 < r_p < 20$ \mpch.
Errors on \wprp \ are calculated using
the standard error across the 10 separate data pointings.  
We estimate the integral constraint numerically using the random
catalog (see e.g., Eqn. 8 of Roche and Eales 1999\nocite{Roche99}) and find
it to be $\sim$0.1 and therefore negligible.

If \xir \ is modeled as a power-law, $\xi(r)=(r/r_0)^{-\gamma}$, then
\rr \ and $\gamma$ can be readily extracted from the projected
correlation function, \wprp, using an analytic solution to Equation
\ref{eqn}:
\begin{equation}
w_p(r_p)=r_p \left(\frac{r_0}{r_p}\right)^\gamma
\frac{\Gamma(\frac{1}{2})\Gamma(\frac{\gamma-1}{2})}{\Gamma(\frac{\gamma}{2})},
\label{powerlawwprp}
\end{equation}
where $\Gamma$ is the gamma function.  A power-law fit to \wprp \ will
then recover \rr \ and $\gamma$ for the real-space correlation
function, \xir.  We correct for the covariance
between $r_p$ bins when calculating the errors on \rr \ and $\gamma$.

To correct for undersampling of galaxies on small scales due to our
slitmask target selection algorithm we use the mock galaxy catalogs of
\cite{Yan03}.  We create samples with identical redshift and
luminosity ranges as each data sample and measure the ratio of the
projected correlation function in the mock catalogs for catalogs with
and without the slitmask target selection algorithm applied.  We then
multiply the measured \wprp \ in the data by this ratio, which is a
smooth function of both scale and luminosity.  We note that using mock
catalogs with different HOD parameters does not change our results
\citep{Coil05}. Both the observed and corrected data are shown in this
paper.

\section{Clustering as a Function of Luminosity}

\begin{figure*}[t]
\centerline{\scalebox{0.5}{\includegraphics{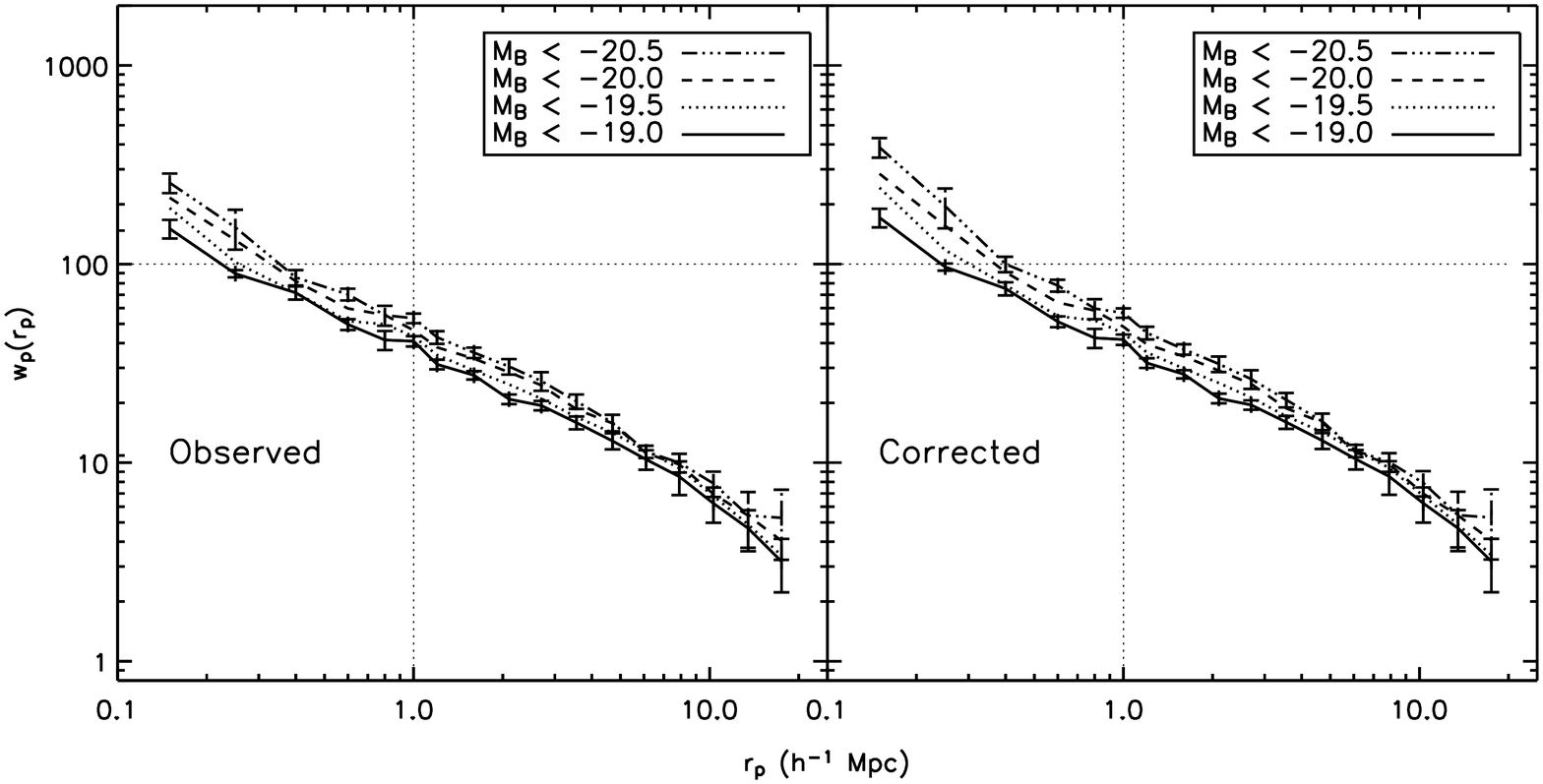}}}
\caption{Projected two-point correlation function, \wprp, of DEEP2
  galaxies for different luminosity samples.  The observed \wprp \ is
  shown in the left panel, while on the right we have corrected for
  the slitmask target selection effects using mock catalogs.  Fiducial
  lines at $r_p=1$ \mpch \ and \wprp$=100$ have been drawn to guide the eye.
\label{lumplot}}
\end{figure*}

\begin{figure}[t]
\centerline{\scalebox{0.35}{\includegraphics{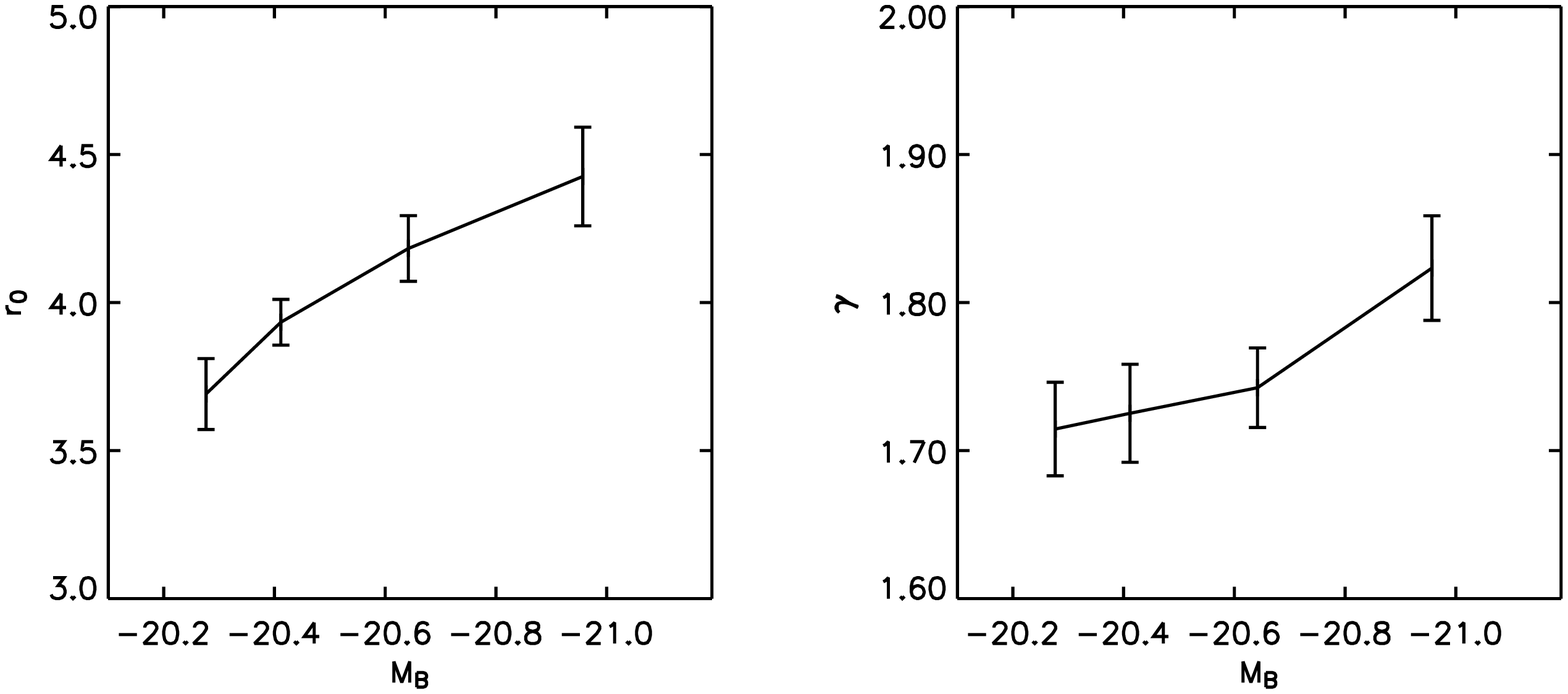}}}
\caption{The clustering scale-length, \rr \ (left), and slope, $\gamma$
  (right), of DEEP2 galaxies as a function of the median absolute
  magnitude of each sample. The values of 
\rr \ and $\gamma$ for each sample are given in Table 2.
\label{r0plot}}
\end{figure}

\begin{figure}[t]
\centerline{\scalebox{0.5}{\includegraphics{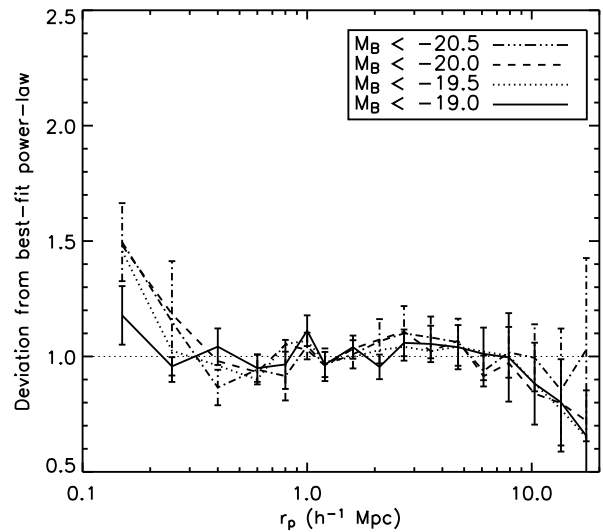}}}
\caption{Deviations of \wprp \ from the best-fit power law for each
  luminosity sample as a function of scale, using the values of \rr \
  and $\gamma$ listed in Table 2.  A thin dotted line is drawn for
  reference at $y=1$.
\label{deviations}}
\end{figure}

Fig. \ref{lumplot} shows the measured projected correlation function
for each luminosity sample as a function of scale, for $r_p=0.1-20$
\mpch.  The left panel shows the observed \wprp, while the right panel
shows \wprp \ after correcting for slitmask target selection effects.
The corrections are larger for the brighter samples and are most
significant on scales less than $r_p=0.3$ \mpch.  Errors are
calculated using the standard error across the ten individual DEEP2
pointings.  Amplitudes and errors for \wprp \ are given for each of the
luminosity samples at four different scales in Table \ref{amptable}.

Power-law fits to the corrected \wprp \ are performed for a range of scales: 
the full range shown here ($0.1<r_p<20$ \mpch), smaller scales 
($0.1<r_p<1$ \mpch), and larger scales ($1<r_p<20$ \mpch). 
Values of \rr \ and $\gamma$ are given in Table 2 for each $r_p$ range and 
results for the full range are shown in Fig. \ref{r0plot}.  
Errors on \rr \ and $\gamma$
are estimated using jacknife errors across the ten samples; this
approach includes the covariance between adjacent $r_p$ bins and
results in errors which are 2-3 times higher than the formal
$1-\sigma$ statistical errors of the least-squares fit to the data,
assuming the $r_p$ bins are independent.  Fitting for a power law on
larger scales, $1<r_p<20$ \mpch, results in \rr \ and $\gamma$ values
within the quoted 1-$\sigma$ errors for the full range, 
as shown in Table 2, and fits to the observed
data without corrections for slitmask effects result in 2\% lower
values of \rr \ and $\gamma$ for the full range for all samples 
except the brightest, 
where the difference is 3\% in both \rr \ and $\gamma$.

We show in Fig. \ref{deviations} deviations from the best-fit power
law over all scales for each luminosity sample.  A power law 
fits the data reasonably well on scales $r_p>1$ \mpch, but all samples 
depart from
power-law behavior at small scales ($r_p<0.4$ \mpch).  These
deviations can be explained naturally by a halo occupation approach to
modeling the observed clustering; we will present HOD fits to these
data in a separate paper.

The clustering scale length, \rr, is significantly larger for the
brighter samples; \rr \ increases from $3.69 \pm0.14$ to $4.43
\pm0.14$ over the luminosity ranges shown here.  The slope is
$\sim1.73 \pm0.03$ for each of the three samples with $L\leq L^*$ and
increases to $\gamma=1.82 \pm0.03$ for the brightest $L>L^*$ sample
(at $z=1$, $M^*=-20.7$ in our magnitude units; cf. Willmer et
al. 2006\nocite{Willmer06}).  This increase in $\gamma$ is found on
both small ($r_p<1$ \mpch) and large ($r_p>1$ \mpch) scales.  
We note that the ratio of red to blue galaxies is not widely 
different between the samples: 76\% of the galaxies are blue in
the faintest sample while 67\% are blue in the brightest sample (see
Fig. 4 of \cite{Willmer06} for a color-magnitude diagram of DEEP2
galaxies).  

As shown in Fig. 6 of \cite{Conroy05}, the clustering results presented here 
are well-fit by 
a simple model in which luminosities are assigned to dark matter halos
and sub-halos in an N-body simulation by matching the observed DEEP2
luminosity function \citep{Willmer06} to the sub-halo/halo circular velocity
function (as a proxy for mass), with no free parameters.  Every
sub-halo with mass $M > 1.6 \times 10^{10} h^{-1} M_\sun$ (the
resolution limit of the halo catalog) is assumed to have a galaxy at
the center; therefore the radial distribution of galaxies matches that
of the sub-halos. This relatively simple model correctly reproduces
our clustering measurements as a function of luminosity and scale presented
here, on scales $r_p=0.1-10$ \mpch, and implies that brighter
galaxies reside in more massive dark matter halos.

\begin{deluxetable}{ccccc}
\tablewidth{0pt}
\tablecaption{Projected Correlation Function Amplitudes for Luminosity Samples}
\tablehead{
\colhead{}&\multicolumn{4}{c}{$w_p$ at $r_p$ in \mpch} \\
\colhead{}&\colhead{$r_p=0.15$}&\colhead{$r_p=1.0$}&\colhead{$r_p=4.7$}&\colhead{$r_p=10.3$}
}
\startdata
1 & $171 \pm18$ & $41.7 \pm2.5$ & $12.9 \pm1.2$ & $6.2 \pm1.3$ \\
2 & $242 \pm29$ & $44.8 \pm3.3$ & $14.2 \pm1.2$ & $6.8 \pm1.3$ \\
3 & $285 \pm27$ & $48.3 \pm4.2$ & $15.9 \pm0.9$ & $7.0 \pm1.1$ \\
4 & $387 \pm44$ & $56.7 \pm3.1$ & $16.1 \pm1.5$ & $7.9 \pm1.2$ \\
\enddata
\tablecomments{\footnotesize These values have all been corrected for slitmask target selection effects.}
\label{amptable}
\end{deluxetable}

\section{Galaxy Bias}

\begin{figure}[t]
\centerline{\scalebox{0.45}{\includegraphics{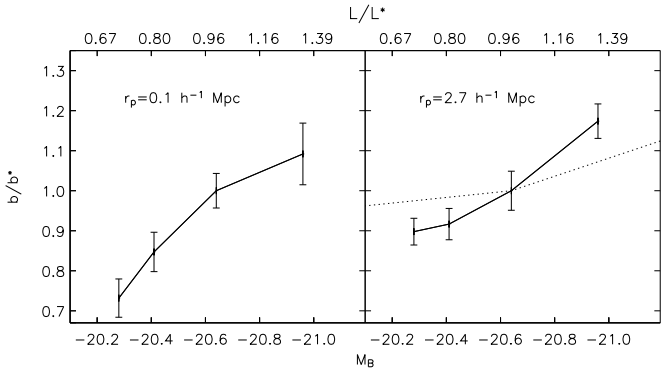}}}
\caption{The mean relative bias of each of our samples compared to the
$M_B<-20.0$ sample.  The relative bias increases linearly with
luminosity, though the trend is stronger on smaller scales (left) than
larger scales (right).  The dotted line in the right panel shows the
same relation from \cite{Zehavi05} in SDSS galaxies at $z\sim0.1$ 
as a function of $L/L^*$, shown in the upper axis.
\label{relbias}}
\end{figure}

We now calculate the relative bias of each luminosity sample with
respect to the $M_B<-20.0$ sample, which has a median luminosity near
$L^*$ ($M^*=-20.7$ at $z=0.9$; see \cite{Willmer06}).  The relative
bias is defined as the square root of the ratio of \wprp \ for a given
sample divided by \wprp \ for the $M_B<-20.0$ sample.  We calculate
the relative bias at two scales, $r_p=0.1$ \mpch \ and $r_p=2.7$
\mpch, using the power-law fits given in Table 2 for small and large
scales. Fig. \ref{relbias} plots the relative bias as a function of
median absolute magnitude (bottom axis) and $L/L^*$ (top axis) at
$r_p=0.1$ \mpch \ (left) and $r_p=2.7$ \mpch \ (right).  There is a
linear trend of bias with luminosity on both small and large scales,
though the trend is stronger on smaller scales.

In the right panel, we also show for comparison the bias of SDSS 
galaxies as a function of $L/L*$ measured by \cite{Zehavi05} on the 
same scale, $r_p=2.7$ \mpch.  The bias of galaxies is a stronger 
function of $L/L^*$ at $z\sim1$ compared to $z\sim0$ over the 
luminosity range that we sample here.

\begin{figure}[t]
\centerline{\scalebox{0.5}{\includegraphics{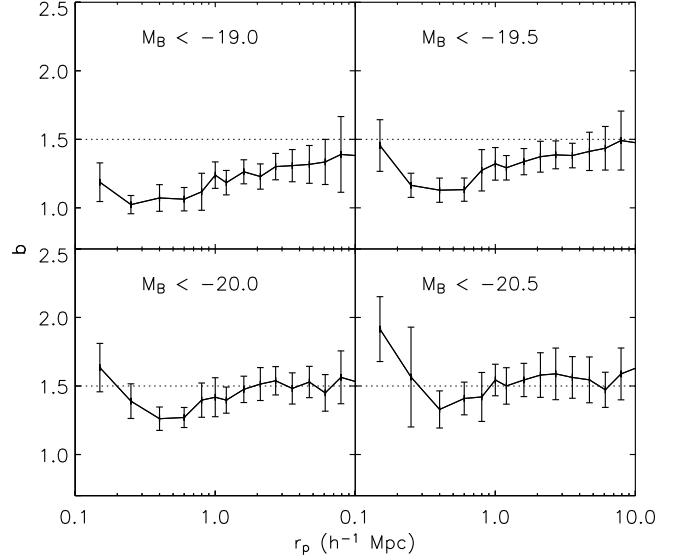}}}
\caption{The absolute bias of DEEP2 galaxies for a $\sigma_8$=0.9
  \lcdm cosmology as a function of scale
for each of our luminosity samples.  The bias is derived using the
ratio of the observed galaxy clustering to the \cite{Smith03} fitting
function for the clustering of dark matter halos at the same
redshifts.  There is a clear scale-dependence to the bias for 
$r_p<1$ \mpch.  The bias on scales $r_p>1$ \mpch \ increases with
  luminosity and is relatively scale-independent.
\label{bias}}
\end{figure}

We further calculate the absolute bias of each sample, using the ratio
of the observed galaxy clustering to the clustering expected for the
underlying dark matter density field.  To estimate the dark matter
correlation function, we use the power spectrum provided by the
publicly available code of \cite{Smith03}, for a cosmology with $\Omega_m=0.3,
\Omega_\Lambda=0.7, \sigma_8=0.9$, and $\Gamma=0.21$.  \cite{Smith03} 
analyze the non-linear evolution of clustering in a large library of 
N-body cosmological simulations, including small scales where 
merging of dark matter halos is important.  They use an analytic 
halo-model approach to fit the non-linear structure evolution, 
which is more accurate than the popular \cite{Peacock94} prescription; 
we include a 5\% error on
the dark matter \wprp \ in our calculation of the absolute bias to
reflect the uncertainty in the fit.  We do not include an additional
uncertainty in the cosmological parameters; for $\sigma_8=0.8$ the
large-scale bias is 13\% higher, while for $\sigma_8=1.0$ it is 10\%
lower.  The results are shown in Fig. \ref{bias}.  While there is no
significant scale-dependence to the bias on scales $r_p>1$ \mpch, all
four samples show a dip in the bias on scales $r_p\sim0.3-0.8$ \mpch \
and a rise on small scales below $r_p\sim0.1$ \mpch.
On scales of $r_p=1-10$ \mpch, the mean absolute bias ranges from
$b=1.26 \pm0.04$ for the $M_B<-19$ sample to $b=1.54 \pm0.05$ for the
$M_B<-20.5$ sample; values are listed in Table 2. On the smallest
scales we measure here, $r_p=0.1$ \mpch, the bias ranges from $b=1.2
\pm0.1$ for the $M_B<-19$ sample to $b=1.9 \pm0.2$ for the $M_B<-20.5$
sample.

We note that the absolute biases measured here are larger than in
\cite{Coil03xisp} (though within 1$\sigma$, given the error on that result), 
where we measured the clustering of all DEEP2
galaxies in the first completed pointing.  This is simply due to
cosmic variance: galaxies in that pointing exhibit lower clustering
than the mean of the completed DEEP2 survey; that region happened to
be less clustered than average.

From these large-scale bias measurements we can estimate the minimum
dark matter halo mass that each galaxy sample resides in, on average,
where the mass is defined as the mass enclosed in a region that is 200
times the critical density.  Using the formulae of \cite{Sheth99} for
$\Omega_m=0.3, \Omega_\Lambda=0.7$, and $\sigma_8=0.9$, we find that
the minimum dark matter mass for our samples ranges from $\sim9
\times10^{11}-3 \times10^{12} h^{-1} {\rm M}_\sun$ (values are listed
in Table 2).

We compare our results with those of the Vimos-VLT Deep Survey (VVDS).
 \cite{Marinoni05} measure the luminosity-dependence of the galaxy bias
averaged on scales 5-10 \mpch \ using 
a sample 1044 VVDS galaxies in an area of $0.16$ square degrees between
$0.7<z<0.9$.  They find that the bias is 
$1.02 \pm0.20$ and $1.14 \pm0.21$ for luminosity thresholds of 
$M_B<-18.7$ and $M_B<-20$, where the quoted errors include the
statistical error only and do not include the dominant error 
due to cosmic variance.  Their bias values are lower than those shown 
here but are likely within the errors if cosmic variance is included.
Additionally, a more recent VVDS paper on the luminosity-dependence 
of galaxy clustering at $0.5<z<1.2$ \citep{Pollo06} yields correlation 
amplitudes and slopes that are similar to ours, but with a steeper trend 
in $b/b*$; they find lower $r_0$ values for the fainter samples 
($r_0=2.95 \pm0.34$ for $M_B<-19$) and higher $r_0$ values for the 
brighter samples ($r_0=5.01 \pm1.5$ for $M_B<-21$) and 
a stronger change in correlation slope $\gamma$ with luminosity than 
is found here.  However, given the cosmic variance errors in the results, 
which use a sample of $\sim$3300 galaxies with $M_B<-19$ covering an 
area of $0.49$ square degrees, the differences are not significant.

\begin{deluxetable*}{ccccccccc}
\tabletypesize{\footnotesize}
\tablewidth{0pt}
\tablecaption{Clustering Results for Luminosity Samples}
\tablehead{
\colhead{}&\colhead{$r_0$}\tablenotemark{a}&\colhead{$\gamma$} &\colhead{$r_0$}\tablenotemark{b}&\colhead{$\gamma$} &\colhead{$r_0$}\tablenotemark{c}&\colhead{$\gamma$}&\colhead{b\tablenotemark{d}}&\colhead{${\rm M}_{min}$}  \\
\colhead{}  &\colhead{\mpch}&\colhead{}         &\colhead{\mpch}&\colhead{}         &\colhead{\mpch}&\colhead{}        &\colhead{}&\colhead{$ h^{-1} {\rm M}_\sun$} 
}
\startdata
1   & $3.69 \pm0.14$ & $1.71 \pm0.03$ & $3.78 \pm0.23$ & $1.68 \pm0.04$ & $3.76 \pm0.09$ & $1.76 \pm0.05$ & $1.26 \pm0.04$ & $9.0 \ 10^{11}$ \\
2   & $3.93 \pm0.10$ & $1.73 \pm0.03$ & $3.73 \pm0.19$ & $1.80 \pm0.05$ & $3.97 \pm0.10$ & $1.74 \pm0.06$ & $1.36 \pm0.04$ & $1.4 \ 10^{12}$ \\
3   & $4.18 \pm0.08$ & $1.74 \pm0.03$ & $3.67 \pm0.11$ & $1.93 \pm0.06$ & $4.25 \pm0.12$ & $1.76 \pm0.07$ & $1.48 \pm0.04$ & $2.2 \ 10^{12}$ \\
4   & $4.43 \pm0.14$ & $1.82 \pm0.03$ & $4.21 \pm0.22$ & $1.90 \pm0.09$ & $4.47 \pm0.11$ & $1.84 \pm0.05$ & $1.54 \pm0.05$ & $2.8 \ 10^{12}$ \\
\enddata
\tablenotetext{a}{for scales $r_p=0.1-20$ \mpch}
\tablenotetext{b}{for scales $r_p=0.1-1$ \mpch}
\tablenotetext{c}{for scales $r_p=1-20$ \mpch}
\tablenotetext{d}{large-scale bias ($r_p=1-10$ \mpch)}
\label{restable}
\end{deluxetable*}

\section{Conclusions}

Using volume-limited subsamples of the 
25,000 galaxies with spectroscopic redshifts between $0.7\leq z
\leq1.4$ from the nearly-completed DEEP2 Galaxy Redshift Survey, 
we have measured the clustering of galaxies as a function of
luminosity at $z\sim1$.  We find that the clustering scale length,
\rr, increases linearly with luminosity over the range sampled here and
that the bias of clustering relative to $L^*$ is strong to what is found 
at $z\sim0$.  The brightest DEEP2 sample, with $L>L*$, has a
significantly steeper correlation slope ($\gamma=1.82 \pm0.03$) on all
scales than the fainter luminosity samples, which all have similar
slopes of $\gamma\sim1.73$.  This is not due to differences in the mix
of galaxies in each sample; the red galaxy fraction depends only
weakly on luminosity in the DEEP2 data.

\cite{Conroy05} attempt to model the clustering results presented here with
a simple prescription in which luminosities are assigned to dark matter halos
and sub-halos in an N-body simulation by matching the observed DEEP2
luminosity function to the sub-halo/halo maximum circular velocity, used as
a proxy for mass.  Their model fits our data extremely well, implying that 
brighter galaxies reside in more massive dark matter halos and that the
luminosity-dependent clustering is dominated by the physical
distribution of halos on large scales and sub-halos on small scales,
as a function of mass.  It is remarkable that the results 
presented in this paper can be explained 
using a model in which the luminosity of a galaxy depends solely on
the mass of the dark matter halo in which it resides.

The mean galaxy bias in the DEEP2 data on scales $r_p=1-10$ \mpch \
for a concordance cosmology ranges from $1.26 \pm0.04$ to $1.54
\pm0.05$ as a function of luminosity for $L/L^*=0.7-1.3$, 
and there is a scale-dependence to the bias for $r_p<1$ \mpch.  
The brightest samples show the
strongest rise in the correlation function on small scales, which is
likely due to these galaxies preferentially residing in groups.
This upturn on small scales has also been seen both locally for SDSS
galaxies with $L>L^*$ \citep{Zehavi05} and at high redshift for LBG
galaxies at $z\sim3-4$ \citep{Lee05,Ouchi05}.  The strength of the
rise in the DEEP2 sample is smaller than what is found at $z\sim3-4$,
as expected from simulations \citep{Kravtsov04}, and the large-scale
bias is intermediate between the bias of LBGs and local galaxies of
similar $L/L^*$ in 2dF and SDSS.

\acknowledgements 

We thank Scott Burles, 
Charlie Conroy, Darren Croton, Daniel Eisenstein and Zheng Zheng for 
helpful discussions.  We also thank the referee for useful comments.  
This project was supported by the NSF grant
AST-0071048. A.L.C. and J.A.N. are supported by NASA through Hubble
Fellowship grants HF-01182.01-A and HST-HF-01165.01-A, awarded
by the Space Telescope Science Institute, which is operated by the
Association of Universities for Research in Astronomy, Inc., for NASA,
under contract NAS 5-26555.  
The DEIMOS spectrograph was funded by a grant from CARA (Keck
Observatory), an NSF Facilities and Infrastructure grant (AST92-2540),
the Center for Particle Astrophysics and by gifts from Sun
Microsystems and the Quantum Corporation. 
The data presented herein were obtained at the W.M. Keck Observatory, which is
operated as a scientific partnership among the California Institute of
Technology, the University of California and the National Aeronautics
and Space Administration. The Observatory was made possible by the
generous financial support of the W.M. Keck Foundation. The DEEP2 team
and Keck Observatory acknowledge the very significant cultural role
and reverence that the summit of Mauna Kea has always had within the
indigenous Hawaiian community and appreciate the opportunity to
conduct observations from this mountain.

\end{document}